\documentclass[5p]{elsarticle}

\usepackage[utf8]{inputenc}
\usepackage[T1]{fontenc}

\usepackage{siunitx}
\DeclareSIUnit\fm{fm}
\DeclareSIUnit\feV{feV}
\DeclareSIUnit\aeV{aeV}

\usepackage{dblfloatfix}
\usepackage{graphicx}
\graphicspath{{figures/}}
\usepackage{mhchem}
\usepackage{isotope}
\usepackage{fixme}
\usepackage[font=footnotesize,labelfont=bf]{caption}
\usepackage[font=footnotesize,labelfont=bf]{subcaption}
\usepackage{lineno,hyperref}
\usepackage[capitalise]{cleveref}
\usepackage{booktabs}

\journal{Physics Letters B}

\newcommand{\Li}{\isotope[7]{Li}}
\newcommand{\Be}{\isotope[8]{Be}}
\newcommand{\Boron}{\isotope[8]{B}}

\newcommand{\ndf}{\text{ndf}}

\newcommand{\mauth}{\author}
\begin{document}

\begin{frontmatter}

  \title{Measurement of the full excitation spectrum of the \ce{^7Li}(p,$\gamma$)$\alpha \alpha$ reaction at \SI{441}{\keV} }

\mauth{Michael Munch}
\cortext[mycorrespondingauthor]{Corresponding author}
\ead{munch@phys.au.dk}
\mauth{Oliver Sølund Kirsebom}
\mauth{Jacobus Andreas Swartz}
\mauth{Karsten Riisager}
\mauth{Hans Otto Uldall Fynbo}

\address{Department of Physics and Astronomy, Aarhus University, Denmark}

\begin{abstract}
A current challenge for ab initio calculations is systems that contain large continuum
contributions such as \Be. We report on new measurements of radiative decay widths in this
nucleus that test recent Green's function Monte Carlo calculations.

Traditionally, $\gamma$ ray detectors have been utilized to measure the high energy photons from the
$\Li(p,\gamma)\alpha \alpha$ reaction. However, due to the complicated response function of these detectors it
has not yet been possible to extract the full $\gamma$ ray spectrum from this reaction. Here we
present an alternative measurement using large area Silicon detectors to detect the two $\alpha$
particles, which provides a
practically background free spectrum and retains good energy resolution.

The resulting spectrum is analyzed using a many-level multi channel R-matrix
parametrization. Improved values for the radiative widths are extracted from the R-matrix fit.
We find evidence for significant non-resonant continuum contributions and
tentative evidence for a broad $0^{+}$ resonance at \SI{12}{\MeV}.
\end{abstract}

\end{frontmatter}

\section{Introduction}
\label{sec:introduction}

In recent years \emph{ab initio} calculations of atomic nuclei,
such as Green's function Monte Carlo (GFMC)~\cite{Pastore2014} and No
Core Shell Model (NCSM)~\cite{Baroni2013}, have advanced tremendously
and now provide quite accurate predictions for light
nuclei. Historically, NCSM has struggled with highly clustered states.
However,
the method has recently been combined with the resonating group method
(RGM) to better describe clustered nuclei including continuum properties
\cite{Baroni2013}.

In this context \Be{} provides an interesting benchmark. All states in
this isotope are unbound with its ground state located just
\SI{92}{\keV} above the 2$\alpha$ threshold.
The lowest two states are highly clustered while some of the resonances at higher energy couple
relatively weakly to the 2$\alpha$ final state.

GFMC calculations of electromagnetic
transitions in \Be{} have been performed by Pastore \textit{et al.}~\cite{Pastore2014}, and
experimentally $\gamma$ decays of several states in \Be{} have been
measured.  The focus of the present letter is the $\gamma$ decay of
the \SI{17.64}{\MeV} $1^{+}$ state. M1 decays of this state could
populate both 0$^+$ and 2$^+$ states. There
are two measurements of the transition strength to the ground- 
and first excited states in \Be~\cite{Fowler1949,Zahnow1995}, and two
measurements of transitions to the 2$^+$ doublet at 16.6-16.9MeV
\cite{Paul1968,Sweeney1969}. However, due to the complicated response function of previous
measurements it has not been possible to extract the full $\gamma$ ray spectrum - specifically none
of the previous measurements were
sensitive to $\gamma$ decays into the unresolved energy region
below the 2$^+$ doublet.

This region was resolved experimentally using e.g. $\alpha$-$\alpha$ scattering and the
$\beta$-decay of \Boron{} and \isotope[8]{Li}~\cite{Warburton1986}. To understand these
different ways of populating \Be{}, it is necessary to have
contributions not only from the known resonances, but also a broad
contribution~\cite{Warburton1986} between the first excited state at 3
MeV and the isospin mixed 2$^+$ doublet at 16.6-\SI{16.9}{\MeV}. It is unclear
if this contribution represents a 2$^+$ intruder state, a
non-resonant continuum contribution, or the low energy tails of high
energy resonances~\cite{Warburton1986,Tilley2004}. From theory there
is also a prediction of a 0$^+$ T=0 intruder state at around 12 MeV
\cite{Caurier2001}.

In this letter we will present a measurement of the $\gamma$ decay of
the \SI{17.64}{\MeV} $1^{+}$ state using a method which is sensitive
to this region of interest and essentially background free. By this
method we will not only address the question of intruder states, but
also derive new more reliable values for the partial decay widths of
the already measured transitions.

It should be noted that electromagnetic transitions from the 1$^+$
states of \Be{} are also of high current interest due to the observation of
anomalous internal pair creation in \Be{} and the interpretation of that
as a possible indication of a new light, neutral
boson~\cite{Krasznahorkay2016,Kozaczuk2017}.
\section{Experiment}
\label{sec:experiment}

\begin{figure}[!b]
  \centering
  \includegraphics[width=\columnwidth]{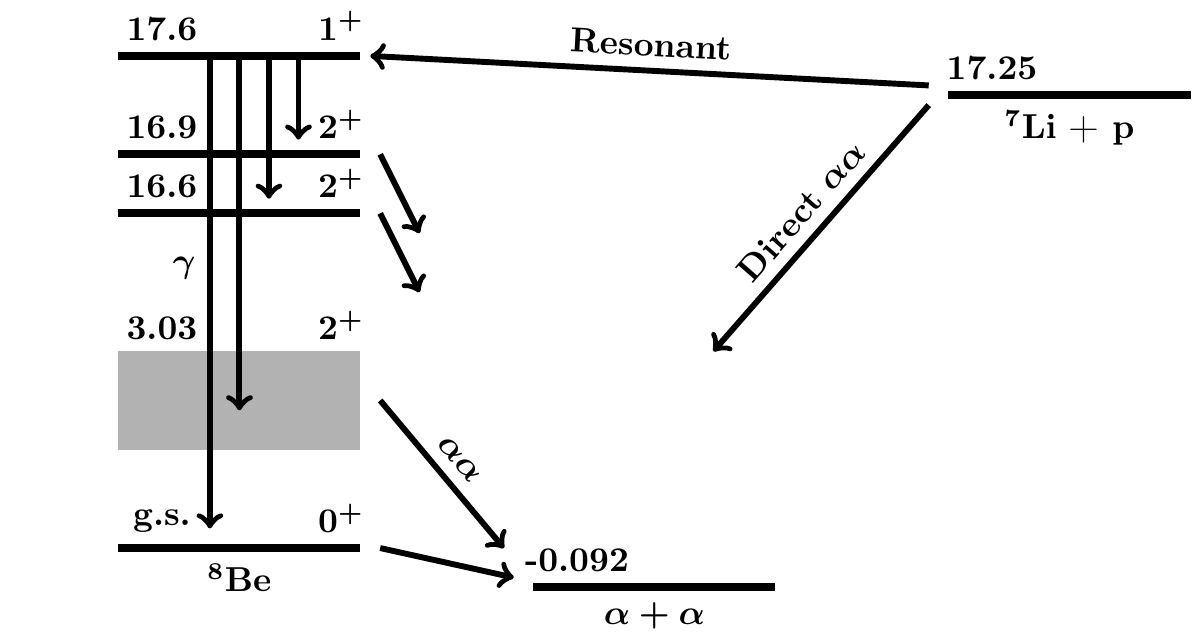}
  \caption{Decay scheme. Only levels populated in the $p + \Li$
    reaction or the $\gamma$
    subsequent decay is shown. Energies are in MeV relative to the \Be{} ground state.}
  \label{fig:scheme}
\end{figure}

The experiment was conducted at the 5MV Van de Graaff accelerator at
Aarhus University that provided a beam of \ce{H_3^+} with energies
between \SI{1305}{\keV} and \SI{1410}{\keV}. The \SI{17.64}{\MeV}
state was populated using the $\Li$(p,$\gamma$) reaction as
illustrated on \cref{fig:scheme}. The beam current was measured using
a suppressed Faraday cup \SI{1}{\m} downstream of the target. Typical
beam currents were between \SI{200}{\pA} and \SI{1}{\nA} and the beam
spot was defined by a pair of $1\times1$ \si{mm} vertical and
horizontal slits. The beam impinged on a natural LiF target
manufactured in house by evaporation of a \SI{160}{\nm} ($\pm 10\%$)
layer of natural lithium fluoride onto a thin
$\sim \SI{4}{\ug\per\cm\squared}$ carbon backing.

The \SI{17.64}{\MeV} state was populated resonantly via $\Li(p,\gamma)$,
as depicted in Fig. 1. While gamma rays were not directly observed, the occurrence of
electromagnetic de-excitation was inferred indirectly from the energies of the two $\alpha$
particles emitted in the subsequent breakup. Charged particles were detected
with two double-sided silicon strip detectors (DSSD) of the W1 type from Micron Semiconductors
\cite{Tengblad2004} 
 giving a simultaneous measurement of position and energy.
 Each detector had an active area of $5\times5$ \si{cm} divided into
$16\times16$ orthogonal strips and was positioned \SI{4}{\cm} from the
target at \SI{90}{\deg} with respect to the beam axis.

A resonance scan was performed with proton energies from 435 to
\SI{470}{\keV} and afterwards data was acquired at \SI{446}{\keV} for
52 hours and at \SI{455}{\keV} for 63 hours.

\begin{figure}[t]
  \centering
  \includegraphics[width=\columnwidth]{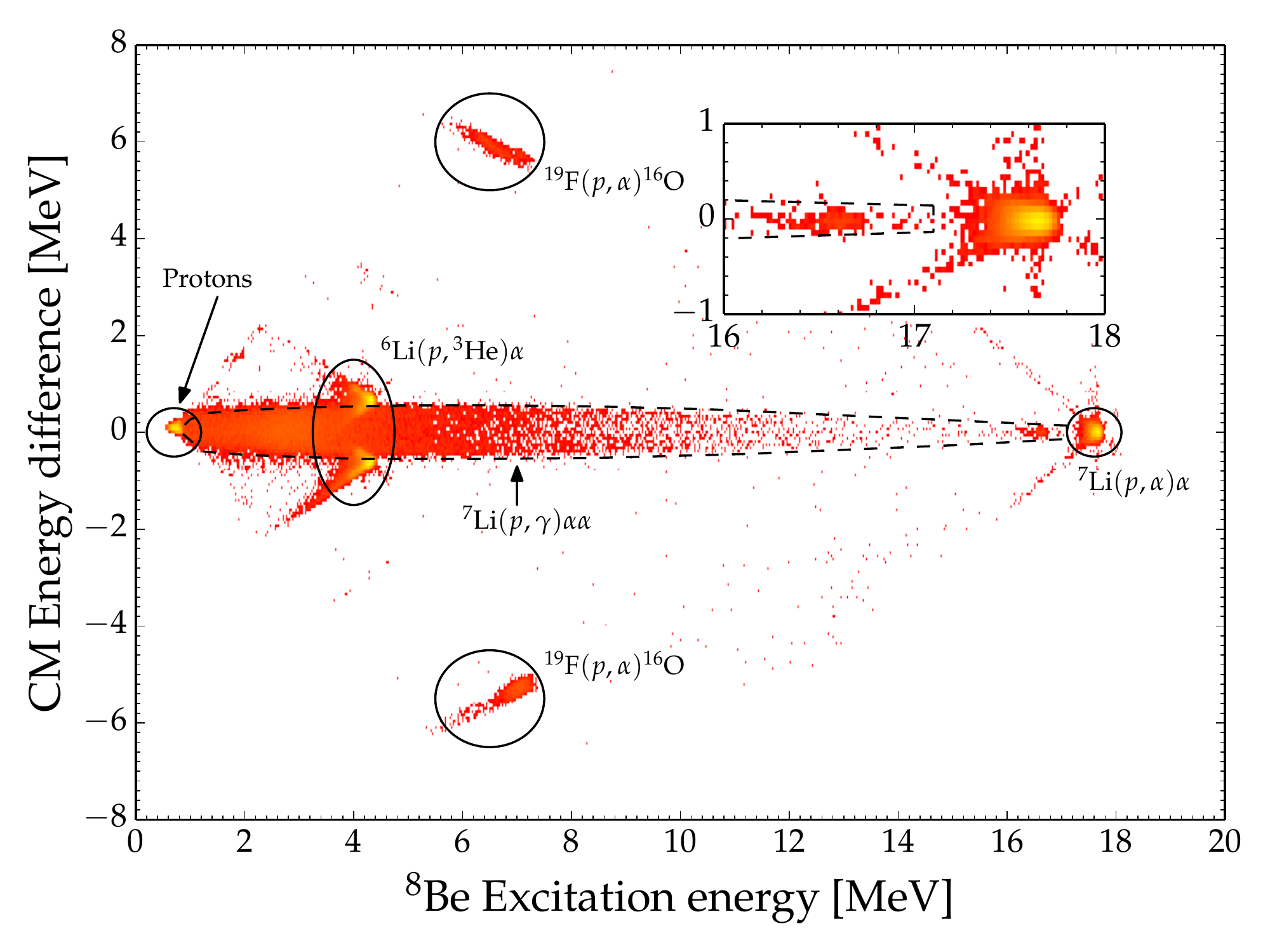}
  \caption{Difference in CM energy vs \Be{} excitation energy. The circles mark various
    background reactions while the band within the dashed contour stretching from 1 to
    \SI{17}{\MeV} corresponds to $\gamma$
    delayed $\alpha$ particles. The insert shows the high excitation energy region. The color scale
  is logarithmic.}
  \label{fig:E0E1cm-ex}
\end{figure}

\begin{figure}[t]
  \centering
  \includegraphics[width=\columnwidth]{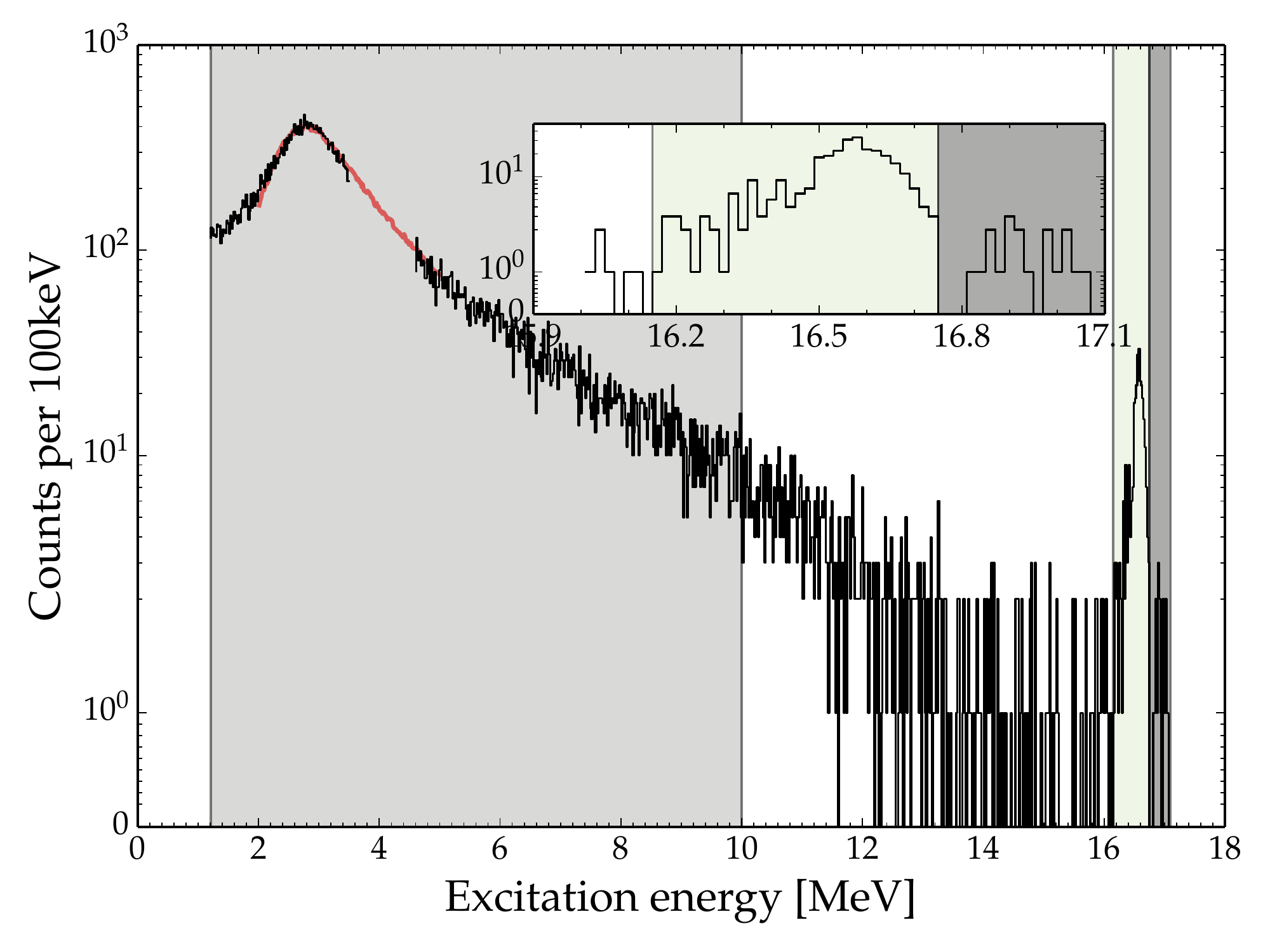}
  \caption{Projected excitation spectrum. The superimposed curve is the best fit to the peak of
    first excited state with a single level R-matrix formula. See \cref{sec:widths} for
    details.}
  \label{fig:ex}
\end{figure}

\section{Data reduction}

The data was analyzed using the full kinematic approach as described in
Ref. \cite{Alcorta2009}. The signal of interest is two coincident $\alpha$ particles with missing
energy corresponding to the reaction $p + \Li \rightarrow \Be^{*} \rightarrow \gamma + \alpha + \alpha$ as illustrated in \cref{fig:scheme}. 

Our coincidence requirement is a time difference of less than \SI{13}{\ns}. As our coincidence
timing resolution is \SI{9.3}{\ns} FWHM this includes $>99\%$
of all true coincidences.  All coincidences surviving this cut are then corrected for energy
loss in the detector dead-layer and target foil assuming they were $\alpha$
particles. The energy of each particle in the center of mass (CM) of $p+\Li$
reaction was determined from its direction and energy. With a simultaneous detection of two
$\alpha$
particles one can infer the corresponding \Be{}
excitation energy from their summed 4-momentum.  \Cref{fig:E0E1cm-ex} shows the difference in
CM energy versus the \Be{}
excitation energy.
In the limit of zero recoil, conservation of energy and momentum dictates that the two alpha
particles should have equal CM energies. When the small, but finite, recoil is taken into
account, the CM energy-difference distribution remains centered very close to zero, but
acquires a sizable spread. Hence the
horizontal band in the figure corresponds to the $\Li(p,\gamma)\alpha \alpha$
reaction.  At high excitation energy there is a distinct peak corresponding to the direct reaction
$\Li(p,\alpha)\alpha$. The two weak diagonal bands extending from the peak correspond to events with
insufficient energy loss correction. These do not interfere with the region of interest and
their strength is negligible compared to the peak.
There are two similar peaks at roughly \SI{4}{\MeV}, which both correspond to
$\ce{^{6}Li}(p,\alpha)\ce{^{3}He}$.
At \SI{7}{\MeV} there are two bands with large deviations from equal energy. This is a
background reaction on fluorine $\ce{^{19}F}(p,\alpha)\ce{^{16}O}$.
At low energy we see random coincidences with the beam.  The identity of the various components
was verified with a Monte Carlo simulation. The $\alpha$-source
energy calibration of the excitation spectrum was cross checked against the
$\ce{^{6}Li}(p,\alpha)\ce{^{3}He}$
and $\Li(p,\alpha)\alpha$
peaks and was found to agree within \SI{4}{\keV} with the tabulated values \cite{Tilley2004}.
It should be stressed that this spectrum is essentially background free in the region of
interest, except for the small region around the $\ce{^{6}Li}(p,\alpha)\ce{^{3}He}$
peaks, which will be excluded from the further analysis.

In order to completely remove random coincidences with the beam we require the  angle between a pair to be
$> \SI{170}{\degree}$ and place a low energy cut at \SI{1}{\MeV}. These cuts preserve
\SI{99}{\%} of the good events. The events corresponding to $\gamma$ delayed $\alpha$ emission are
selected as those within the dashed contour seen on \cref{fig:E0E1cm-ex}.

\Cref{fig:ex} shows the projected excitation spectrum with the first excited state visible at
\SI{3}{\MeV} and the two contributions from the doublet at high energy in the insert. 
The superimposed curve will be discussed in \cref{sec:widths}. The extracted excitation
spectrum can be found in \cite{Dataset}.

\subsection{Normalization}
\label{sec:normalization}

\begin{figure}[t]
  \centering
  \includegraphics[width=\columnwidth]{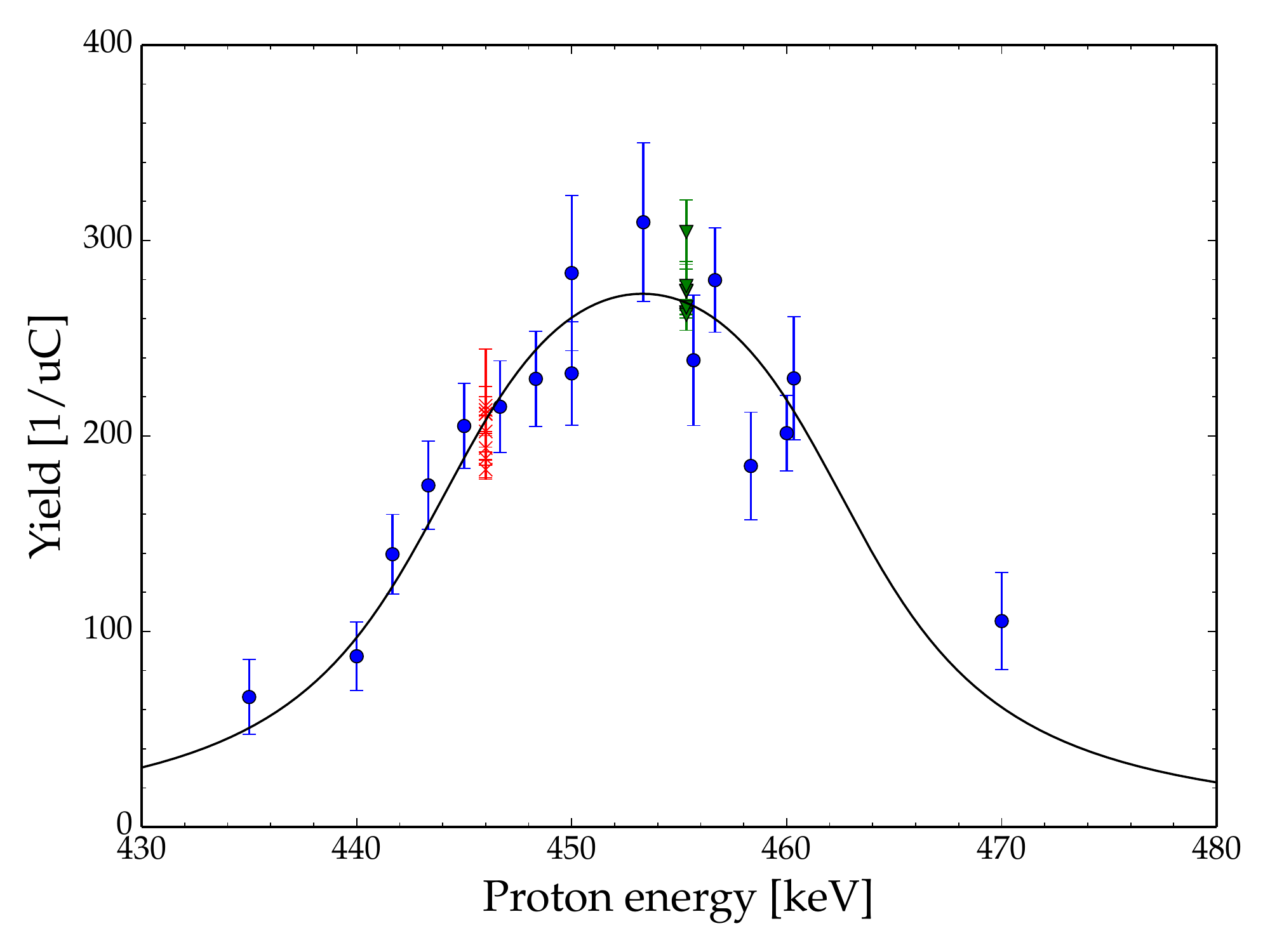}
  \caption{Resonance scan showing the yield of events with an excitation energy between 2 and
    \SI{3}{\MeV} as a function of beam energy. The red crosses and green triangles correspond
    to the long measurements. 
    Each datapoint corresponds to a run and thus slightly different accelerator settings.
    The curve is the best fit to \cref{eq:scan-yield} - see text for details.}
  \label{fig:scan}
\end{figure}

\Cref{fig:scan} shows the yield of events with an excitation energy between 2 and
\SI{3}{\MeV}. The red crosses and green triangles correspond to the two long measurements. The solid line shows the
best fit to equation 14 from Ref. \cite{Fowler1948}. 
\begin{align}
  \label{eq:scan-yield}
  Y = \Big[\tan^{-1} \frac{E_{p}-E_{r}}{\Gamma_{\mathrm{lab}}/2} &- \tan^{-1} \frac{E_{p}-E_{r}-\Delta
                                                E}{\Gamma_{\mathrm{lab}}/2}\Big] \notag \\
                                              &\times \frac{2\pi}{k_{r}^{2}} \frac{g_{J}}{\epsilon} \Gamma_{\gamma},  
\end{align}
where $\Gamma_{\mathrm{lab}}$
is the resonance width in the lab system, $E_{p}$
is the beam energy, $E_{r}$
the resonance energy, $\Delta E$
the energy loss through the target, $g_{J}$ the statistical factor from spin coupling, $k_{r}$
the \Li-p wave number at the resonance energy, and $\epsilon = \frac{1}{N}\frac{dE}{dx}$, where $N$ is the
number density of target nuclei and $\frac{dE}{dx}$ the stopping power.

$\Gamma_{\mathrm{lab}}$
was fixed to $8/7$ of the literature value of $\SI{10.7(5)}{\keV}$ \cite{Tilley2004}.
The last part of the equation was treated as a scaling constant and fitted.  The best fit was
achieved with $\Delta E = \SI{18.1(16)}{\keV}$
and $E_{r} = \SI{444.3(6)}{\keV}$.
The resonance energy is slightly higher than the latest literature value of \SI{441.4(5)}{\keV}
\cite{Tilley2004}.

Upon impinging on the target foil the $\ce{H_{3}^{+}}$
molecule will break up. In this process additional electron stripping, neutralization and
scattering outside the Faraday cup might occur. The effect of these processes can be determined
by measuring the integrated current with and without target foil placed in the beam. The ratio
of these two measurements gives the effective charge state of the $\ce{H_{3}^{+}}$
molecule as observed at the Faraday cup, when the beam passes through the foil. The result was
$\num{2.50(7)}e$ over the measured energy range.

\section{Extraction of radiative widths}
\label{sec:extr-radi-widths}

As previous experiments have determined the widths using simple
integration of the excitation spectrum, we will first determine the widths using this
method. In addition, we will perform an R-matrix analysis of the measured spectrum in order to
take interference into account. The R-matrix parametrization is described elsewhere \cite{Note}. The R-matrix implementation can be found in Ref. \cite{ORM}.

\subsection{Bin integration}
\label{sec:widths}

\begin{table}[b]
  \centering
  \caption{Widths extracted from bin integration of the excitation spectra. Literature values
    are from Ref.~\cite{Tilley2004}. The GFMC results are from Ref.~\cite{Pastore2014}.
    The R-matrix results are from \cref{sec:r-matrix-analysis}. $\Gamma_{0_{1}}$ of the R-matrix
    results is from model 2, while the rest is from model 3.}
  \label{tab:widths}
  \begin{tabular*}{\columnwidth}{l @{\extracolsep{\fill}} cccc}
    \toprule 
    Parameter & {Present} & {Lit.} & {GFMC} & {R-Mat.}\\
    \midrule
    $\Gamma_{0_{1}}$ (\si{\eV}) & -  & 15.0(18) & 12.0(3) & 13.8(4)\\
    $\Gamma_{2_{1}}$ (\si{\eV})         & 6.0(3)  & 6.7(13)  & 3.8(2) & 5.01(11)\\
    $\Gamma_{2_{2}}$ (\si{\meV})        & 35(3)    & 32(3)    & 29.7(3) & 38(2)\\
    $\Gamma_{2_{3}}$ (\si{\meV})        & 2.1(6)   & 1.3(3)   & 2.20(5) & 1.6(5)\\
    \bottomrule   
  \end{tabular*}
\end{table}

The excitation spectrum, shown in \cref{fig:ex}, has been subdivided into four regions covering
the first excited state from 1 to \SI{10}{\MeV}, the continuum from 10 to \SI{16.1}{\MeV} and
the two doublet states from 16.1 to \SI{16.75}{\MeV} and 16.75 to \SI{17.1}{\MeV}
respectively. The choice of \SI{10}{\MeV} is somewhat abitrary. It is placed sufficiently high
to include the majority of the peak. Superimposed on the data is the best fit between 2 and
\SI{6}{\MeV} to a single level R-matrix expression fed by an M1 decay.  

The widths were determined by integration of the three regions with solid shading. The
contribution from the excluded region was determined from the superimposed R-matrix curve. The
integrals were converted into absolute decay widths using \cref{eq:scan-yield} and the
parameters determined in \cref{sec:normalization}. 

The results and statistical errors are listed in \cref{tab:widths} along with the current
literature values from Ref.~\cite{Tilley2004} and the results of GFMC
 calculations~\cite{Pastore2014}.

\subsection{R-matrix analysis}
\label{sec:r-matrix-analysis}

We will analyze the excitation spectrum using three different models. Model 1 is a
model with one $0^{+}$
ground state and three $2^{+}$
resonances. All states are fed by M1 $\gamma$
decays while the $2_{1}^{+}$
resonance is also fed by E2 decays. All initial values were taken from
Ref. \cite{Tilley2004}. To ensure convergence the energy of the $2_{3}^{+}$,
as well as the $\alpha$
widths of the two highest $2^{+}$
resonances and the ground state, were fixed. Model 2 adds an additional $2^{+}$
level at high energy fed by an M1 transition. All parameters for this additional level were
allowed to vary freely. Furthermore, it was no longer necessary to fix $E_{2_{3}}$. Model 3
adds another $0^{+}$ state. However, in order to achieve convergence it was necessary to fix
the ground state feeding and the position of the $2^{+}$ background pole to the values from
model 2.
The M1 feeding of all $2^{+}$ levels were summed coherently while the M1 contribution to the $0^{+}$
and the E2 feeding were added incoherently. Model 1 has 8 free parameters, model 2 has 12 and
model 3 has 13.

In order to directly compare the experimental spectrum with the spectrum obtained from R-matrix
theory, it is necessary to fold the theoretical spectrum with the experimental response
function. For this experimental setup the response function is well described as a Gaussian
function with an exponential tail - the one tail variant of Ref. \cite{Bortels1987}. The
parameters were determined with a fit to the $\Li(p,\alpha)\alpha$
peak as the effect of the response function is most important for the narrow $2^{+}$
levels above \SI{16}{\MeV}. The best fit was achieved with $\mu=\SI{23.5(5)}{\keV}$,
$\sigma=\SI{27.07(15)}{\keV}$ and $\tau=\SI{39.6(2)}{\keV}$.

\begin{figure}[t]
  \includegraphics[width=\linewidth]{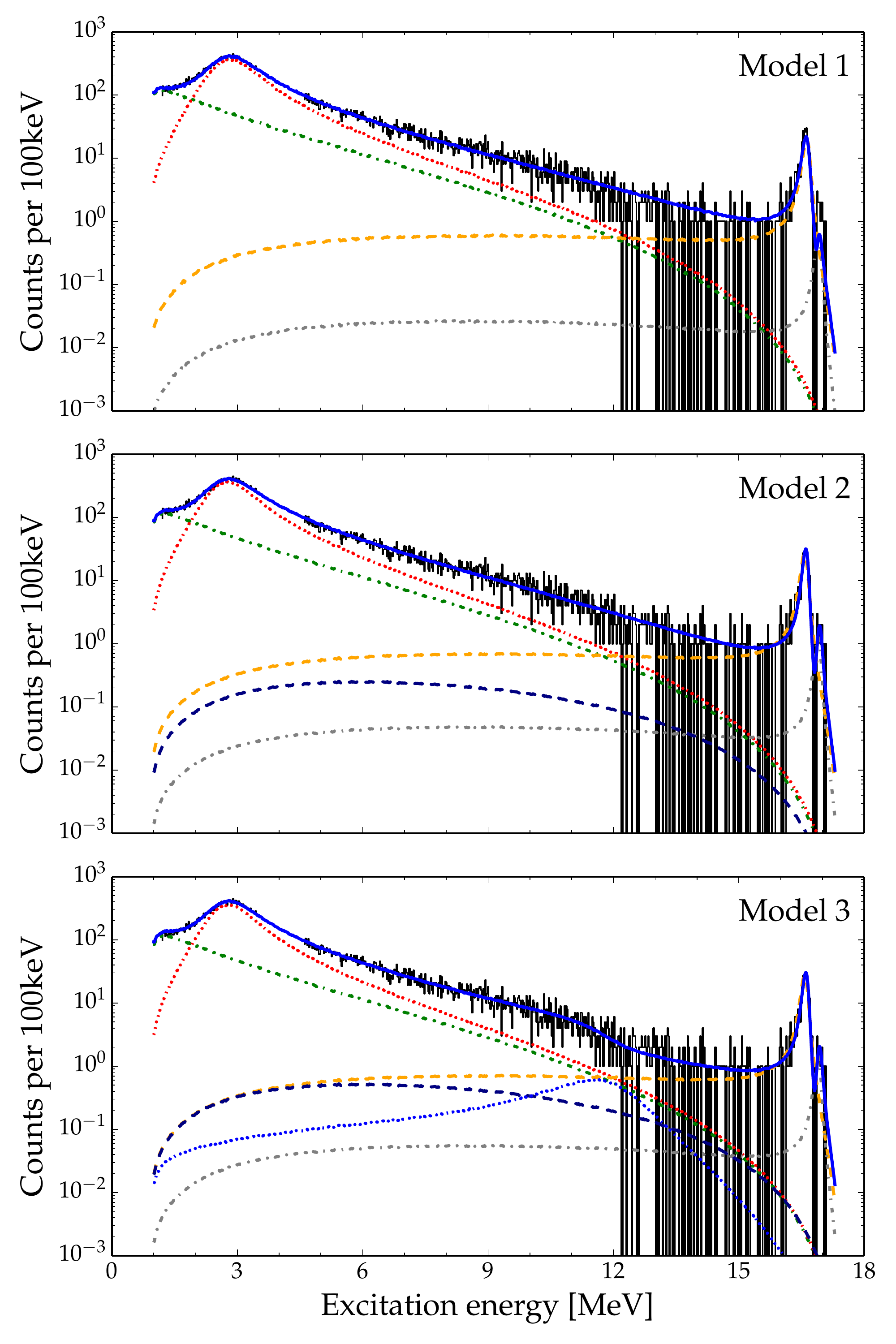}
  \includegraphics[width=\linewidth]{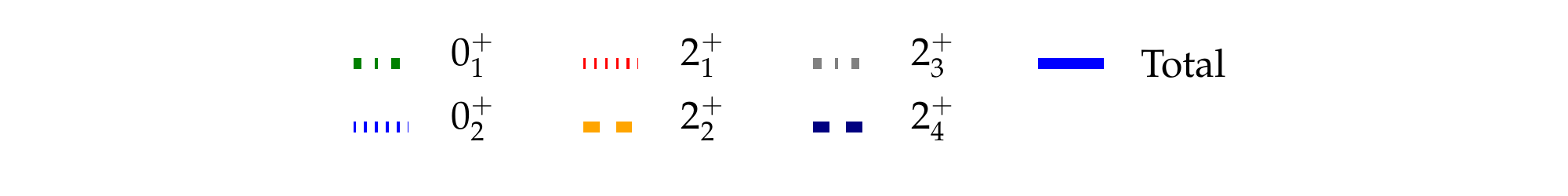}
  \caption{Best fit for a channel radius of \SI{5}{\fm}. The solid blue line shows the sum of all
    contributions, while the dotted shows the single level profile of the individuals levels.}
  \label{fig:fit}
\end{figure}

\Cref{fig:fit} shows the best fit for all models with a channel radius of \SI{5}{\fm}. The dashed
curves show the single level contributions. It should be noted that all models have a weak
dependence on the chosen channel radius, as this influences the shape and peak to tail ratio
for the $0^{+}$ ground state. A channel radius of \SI{5}{\fm} was chosen, as it minimized $\chi^{2}$.

The parameters corresponding to the best fit are listed in \cref{tab:param} along with their
errors. In order to minimize bias a Poisson likelihood estimator has been used
\cite{Bergmann2002}. Errors have been estimated using the MINOS routine and are symmetric unless noted
otherwise. Propagated errors have been calculated using the Hessian approximation.

\begin{table*}[!b]
  \centering
  \caption{Parameters for the best fit for both models with a channel radius of \SI{5}{\fm}.
    Parameters in square brackets were fixed. Decay widths were calculated with eq. (6) from
    the supplementary information. All errors are statistical. Propagated errors are calculated
  using the Hessian approximation.}
  \label{tab:param}
  \sisetup{
    table-space-text-post={)},
    table-space-text-post={()},
    table-align-text-pre = false,
    table-align-text-post = false,
    separate-uncertainty = true
  }
  \newcommand{\fixed}[1]{{[}#1{]}}
  \begin{tabular*}{\textwidth}{l @{\extracolsep{\fill}} c c c}
    \toprule
    Parameter & {Model 1} & {Model 2} & {Model 3}\\
    \midrule
    $E_{0_{1}}$ (keV)                          & \fixed{0} & \fixed{0} & \fixed{0}\\
    $\gamma_{0_{1}M1}$ ($10^{-11}\times\si{\eV^{-1}}$)     & \num{4.35(5)} & \num{4.36(6)} & \fixed{\num{4.36}} \\
    $\Gamma^{0}_{0_{1}M1}$ (\si{\eV})                   & \num{13.7(3)} & \num{13.8(4)} &\fixed{13.8}\\
    $\gamma_{0_{1}\alpha_{0}}$ (\si{\sqrt{\keV}})         & \fixed{22.1} & \fixed{22.1} & \fixed{22.1} \\
    $\Gamma^{0}_{0_{1}\alpha_{0}}$ (\si{\eV})                 & \fixed{5.57} & \fixed{5.57} & \fixed{5.57}\\
    \midrule
    $E_{0_{2}}$ (MeV)                          & - & - & \num{12.0(3)}\\
    $\gamma_{0_{2}M1}$ ($10^{-11}\times\si{\eV^{-1}}$)     & - & - & \num{0.58(8)}\\
    $\Gamma^{0}_{0_{2}M1}$ (\si{\eV})                & - & - & \num{12(3)}\\
    $\gamma_{0_{2}\alpha_{0}}$ (\si{\sqrt{\keV}})         & - & - & \num{-15.2(15)}\\
    $\Gamma^{0}_{0_{2}\alpha_{0}}$ (\si{\MeV})             & - & - & \num{2.4(5)}\\
    \midrule                                                            
    $E_{2_{1},}$ (\si{\keV})                & {$3008\substack{+55\\-9}$} & \num{2960(22)} & \num{2969(11)}\\
    $\gamma_{2_{1},M1}$ ($10^{-11}\times\si{\eV^{-1}}$) & \num{3.31(3)}  & \num{3.22(6)} & \num{3.13(3)}\\
    $\Gamma^{0}_{2_{1},M1}$ (\si{\eV})            & \num{5.57(11)} & \num{5.3(2)} & \num{5.01(11)}\\
    $\gamma_{2_{1},E2}$ ($10^{-22}\times\si{\eV^{-3}}$) & \num{-4.2(12)} & \num{-4(500)} & \num{0.9(592)}  \\
    $\Gamma^{0}_{2_{1},E2}$ (\si{\meV})               & \num{1.9(12)} &  $< \SI{10}{\meV}$ & $< \SI{1}{\meV}$ \\
    $\gamma_{2_{1},\alpha_{2}}$ (\si{\sqrt{\keV}})     &  {$-29.9\substack{+0.3\\-1.5}$} & \num{-29.3(5)} &\num{28.6(3)}\\
    $\Gamma^{0}_{2_{1},\alpha_{2}}$ (\si{\MeV})            &  \num{1701(27)}  & \num{1601(45)} & \num{1546(25)} \\
    \midrule                                                            
    $E_{2_{2}}$ (\si{\keV})                & \num{16629(11)} & \num{16588(5)} & \num{16590(5)}\\
    $\gamma_{2_{2},M1}$ ($10^{-11}\times\si{\eV^{-1}}$) & \num{11.6(7)}   & \num{12.7(4)} & \num{12.9(4)}\\
    $\Gamma^{0}_{2_{2},M1}$ (\si{\meV})              & \num{27.9(17)}  & \num{38(2)} & \num{38(2)}\\
    $\gamma_{2_{2},\alpha_{2}}$ (\si{\sqrt{\keV}})    & \fixed{3.1}     & \fixed{3.1} & \fixed{3.1}      \\
    $\Gamma^{0}_{2_{2},\alpha_{2}}$ (\si{\keV})           & \fixed{108}     & \fixed{108} & \fixed{108}       \\
    \midrule                                                            
    $E_{2_{3}}$ (\si{\keV})                & \fixed{16922}   & \num{16912(25)}  & \num{16910(23)} \\
    $\gamma_{2_{3},M1}$ ($10^{-11}\times\si{\eV^{-1}}$) & {$3.2\substack{+1.7 \\ -0.9}$} & \num{4.3(8)} & \num{4.5(7)}\\
    $\Gamma^{0}_{2_{3},M1}$  (\si{\meV})             & \num{0.8(8)} & \num{1.4(5)} & \num{1.6(5)}\\
    $\gamma_{2_{3},\alpha_{2}}$ (\si{\sqrt{\keV}})     & \fixed{2.2} & \fixed{2.2} & \fixed{2.2}  \\
    $\Gamma^{0}_{2_{3},\alpha_{2}}$ (\si{\keV})            & \fixed{74} & \fixed{74} & \fixed{74} \\
    \midrule                                                            
    $E_{2_{4}}$ (\si{\MeV})                & -   & \num{24(3)}  & \fixed{24}      \\
    $\gamma_{2_{4},M1}$ ($10^{-11}\times\si{\eV^{-1}}$) & - & \num{-1.1(2)} & \num{-1.8(2)}  \\
    $\Gamma^{0}_{2_{4},M1}$  (\si{\meV})             & -            & \num{57(20)} & \num{160(40)}\\
    $\gamma_{2_{4},\alpha_{2}}$ (\si{\sqrt{\keV}})     & -            & \num{38(7)}  & \num{35.9(18)}\\
    $\Gamma^{0}_{2_{4},\alpha_{2}}$ (\si{\MeV})            & -             & \num{20(8)} & \num{18.0(18)}\\
    \midrule
    $\chi^{2}/\ndf$                          & $878/735$ & $838/731$ & $808/730$\\
    $P$ (\%)                          & 0.02 & 0.36 & 2.3 \\
    \bottomrule
  \end{tabular*}
\end{table*}
\section{Discussion}
\label{sec:discussion}

Comparing the radiative widths determined using integration with those from the literature, we
find agreement for the doublet. Previous measurements assigned everything below the ground
state peak in the $\gamma$ spectrum to the $2_{1}$
distribution. In light of this and considering that our measurement of the $2_{1}$
distribution has not been extrapolated to zero energy, the agreement with literature is
reasonable.

The excitation spectrum produced by the three R-matrix models can be seen in \cref{fig:fit} and
in all cases the majority of the experimental spectrum has been reproduced. The main
qualitative improvement between model 1 and 2 is observed around the \SI{16.9}{\MeV} peak,
which model 1 systematically undershoots.
In addition, an order of magnitude improvement of the P-value is observed for each subsequent model.
The single level shape of the model 2 background pole
is a broad featureless distribution. We interpret this as non-resonant continuum contributions.
An interesting aspect of all models is the significant contribution of the ground state in all
energy regions and its dominance below \SI{2}{\MeV}. From this rather remarkable feature it is
possible to determine the ground state strength from a measurement well above the
peak. This behavior is the well known "ghost anomaly" expected
both from theory \cite{Barker1962} and observed in previous experiments on the \Be{} system
\cite{Becchetti1981}. This implies that previous measurements, which
have ignored the anomaly, have overestimated the $2_{1}$
strength by at least 20\%. This estimate is based on the difference between our two
different methods.

Additionally, it should be noted that the observed strength in the intermediate
region between 6 and \SI{16}{\MeV} cannot be attributed to a single resonance but rather a
result of several interfering levels and non-resonant continuum contributions. The extracted $\gamma$
widths for the three models agree within the statistical errors except for the $2_{2}$
width. However, as model 1 systematically deviates in that region we recommend that the model 3
parameters are used. The change in resonance energy is expected from interference and similar
effects were observed in $\beta$ decay experiments \cite{Hyldegaard2010}.

The second $0^{+}$ level, introduced in model 3, interferes destructively with the
ground state, as can be observed in the 12 to \SI{16}{\MeV} region where it improves the
agreement substantially. Its location is interesting as it coincides precisely with the
energy predicted by \cite{Caurier2001}. 

The current literature value of \SI{0.12(5)}{\eV} in Ref. \cite{Tilley2004} for the E2 strength
is based on the width listed in \cref{tab:widths} and a measurement of the E2/M1 mixing ratio
\num{0.018(7)} \cite{Meyer1961}. However, all our models yield a significantly smaller ratio
$< 0.002$.
While the errors involved are too large to draw a conclusion, it is important to note that
there is significant spread in the reported mixing values
\cite{Boyle1956,Grant1960,Mainsbridge1960}. A more detailed measurement of the $\alpha-\gamma$
correlation function could resolve this issue.

Compared with the GFMC calculation in Ref. \cite{Pastore2014} we find poor agreement for the
transitions to the $0^{+}$
and $2_{1}^{+}$
states. For the $2_{1}^{+}$ discrepancy, 
Ref. \cite{Pastore2014} suggests that a lack of continuum contributions could explain this. No
explanation is given for the $0^{+}$
discrepancy. It is interesting to note that a value of $\Gamma_{2_{1}E2} = \SI{0.63(5)}{\eV}$,
as suggested by GFMC, can only be accommodated if the first excited state is made extremely wide
$\sim \SI{1.8}{\MeV}$.
A similar disagreement is observed for the transitions to the doublet. However, the strength of
these transitions depend on isospin mixing of the $1^{+}$
doublet states which was first determined by Barker \cite{Barker1966}. Using slightly different mixing
coefficients resolves some of the issues but create others - see Ref. \cite{Pastore2014} for
details. As these states should be well described in the shell-model, it would be interesting to
compare with NCSM calculations of the transition rates. 

\section{Conclusion}
\label{sec:conclusion}

Coincident $\alpha$
particles from the $\Li(p,\gamma)\alpha \alpha$
reaction at a proton energy of roughly $E_{p} = \SI{441}{\keV}$
have been measured using close geometry silicon strip detectors. This yields a background free excitation spectrum
from 1 to \SI{17}{\MeV}. The $\gamma$
decay widths have been determined using both integration and R-matrix analysis.

The results of the R-matrix analysis show that the ground state contributes significantly to
the full energy range and dominates the spectrum below \SI{2}{\MeV}. This implies that simply
integrating the excitation energy spectrum would overestimate the decay strength to the
first excited state. In order to achieve a good fit to data, it is necessary to include a
$2^{+}$
background pole. This indicates that the spectrum has non-resonant continuum
contributions. Additionally, we find tentative evidence for a broad $0^{+}$
state at \SI{12}{\MeV}. A similar measurement of the \SI{18.1}{\MeV} $1^{+}$ state in \Be{}
could further illuminate this.

The extracted widths for the $2^{+}$ doublet is in agreement with previous measurements, while
the results for the ground and first excited state differ by 8 and \SI{34}{\%} respectively.
A comparison with GFMC calculations shows significant differences between 13 and
\SI{34}{\%}. Determination of the $1^{+}$ isospin mixing might bring clarification.

\section{Acknowledgement}
\label{sec:Acknowledgement}

The authors would like to thank Jonas Refsgaard for his invaluable input on R-matrix
theory. Furthermore, we would like to thank Folmer Lyckegaard for manufacturing the target. We
also acknowledge financial support from the European Research Council under ERC starting grant
LOBENA, No. 307447. OSK acknowledges support from the Villum Foundation.

\section*{References}
\bibliography{IFA019-article}

\end{document}